\documentclass[11pt,twoside]{article}

\usepackage{asp2006}
\usepackage{epsf}
\input{psfig}

\def\HI{H{\,\small I}}
\newcommand{\kms}{$\,$km$\,$s$^{-1}$}

\begin{document}
\title{The interplay between radio jets and ISM in sub-kpc radio sources}   
\author{Raffaella Morganti}   
\affil{Netherland Foundation for Research in Astronomy, Postbus 2,
NL-7990 AA, Dwingeloo \\
 and  \\
 Kapteyn Astronomical Institute, University of Groningen, P.O. Box 800,
9700 AV Groningen, the Netherlands} 

\begin{abstract} 

This paper reviews the evidence for interaction between radio jets and their
environment in small (sub-kpc) and intermediate scale radio sources.
Observations of gas (both neutral hydrogen and ionised gas) have shown the
presence of fast ($> 1000$ km/s) outflows likely originating from this
interaction.  The characteristics of these AGN-driven outflows (e.g.  mass
outflow rate) indicate that they may have a relevant impact on the evolution
of the host galaxy.  We also report on the detection of large \HI\ disks found
around the host galaxies of nearby compact radio sources. Similar structures
have so far not been detected around large radio sources. The presence of
these structures in relation to the evolution of young compact radio sources
is discussed.

\end{abstract}


\section{Interaction between the radio jets and the ISM: why is important?}

In a hierarchical scenario, galaxy mergers play a major role in the formation
of early-type galaxies and they may also provide the way to bring gas to the
central regions, fuel the super-massive black hole (SMBH) and make it
active. However, the hierarchical scenario would not fully explain the
properties of the observed structures unless strong feedback occurs dumping
large amount of energy on the interstellar and intergalactic medium.  The
onset of the quasar activity (including the radio-loud phase) is now
considered as one of the important source of feedback, driving gas out of the
central regions of galaxy, regulating start formation and the growth of the
SMBH \citep{silk98,fab99,mat05,hop05}. Thus, {\sl the initial phase of AGN
can be crucial in the evolution of the host galaxy}.

In the radio-loud phase of the AGN, radio jets and the surrounding
inter-stellar medium (ISM) can interact and affect each others.  If the ISM is
rich and dense, the effect of the surrounding medium on the radio jet can, for
example, momentarily disrupt the jet but it can also frustrate or destroy
it.  On the other hand, the radio jets may have a profound influence on the
medium.  It can produce gas outflows that could clear the circum-nuclear
regions. Thus, the interaction between radio-jets and ISM is one of
the possible mechanisms for triggering outflows (together with radiation
pressure and starburst winds). The radio jets can, therefore, provide one of
the source of feedback.  How relevant these outflows are, is the topic of this
paper.

In radio-loud objects, we know which one are {\sl young AGN}. Based on their
size and on the characteristics of their radio spectrum, these are the
so-called GigaHertz Peaked and the Compact Steep Spectrum (GPS/CSS) sources
(with possibly the High Frequency Peaker being an even young example of radio
sources, see Orienti et al. these Proceedings). A more detailed description of
the characteristics of GPS/CSS is given by Giroletti in these Proceedings.  We
can, therefore, use CSS and GPS radio sources to study their effect on the
surrounding medium.  In particular, we use the kinematics of the gas to trace what is
happening in the central regions of these radio-loud, active galaxies. 
We will focus on atomic neutral hydrogen and ionised gas: these two phases of
the gas are given complementary information and therefore they are both
important in order to obtain a full picture of what is happening.  The
molecular gas has been nicely discussed in Patrick Ogle's talk.

Before starting, let me remind you that gas in the centre of early-type
(field) galaxies - the typical host galaxies of the radio-loud objects
discussed here - is an important and common component.  In general, ionised
gas is detected in the central region of about 70\% of early-type galaxies
\citep[see e.g.][]{sa06}. The kinematics of this gas is often complex, e.g. with
kinematical decoupled cores. This suggests that in most of the cases (although
not in all!) the external origin of the gas is the most likely. For the
neutral hydrogen, we will mainly concentrate on gas detected via \HI\  absorption that for
radio-loud objects allows to explore the small scales, circum-nuclear
regions. 
In radio galaxies, \HI\ absorption detected against their
radio cores and likely associated with their circum-nuclear region is
relatively common.
\HI\ emission is more difficult to be detected at the typical distance of
radio galaxies (but see Sec. 4 for some discussion of \HI\ emission for nearby
radio galaxies).  On the large scale, the \HI\ in emission and the ionised gas
appear to be part of the same structure as they show the same kinematic, often
decoupled from the one of the stars \citep{mo06}.
 
\section{What are the characteristics of the circum-nuclear gas in CSS/GPS 
radio sources?}

In the case of young compact radio sources (CSS/GPS), we would like to answer a number
of questions, among which:

\begin{itemize}

\item Are the jets of these young radio sources really moving through a dense medium?

\item How the characteristics of this gas compare to what found in large
radio galaxies?

\item What is the kinematics of the gas? Do we see evidence of the interaction
between the radio jets and the medium (e.g. outflows)? 
\end{itemize}

\subsection{The atomic neutral hydrogen}
From \HI\ absorption observations, we know that CSS/GPS sources have typically
an high detection rate (compared to extended radio sources), as reported by
\citet{ve03,gup06} - although this does not
appear to be the case for the even smaller (and younger) HFP \citep[see][for
possible explanations]{or06}.  CSS and GPS sources also appear to have often
an higher optical depth (and column density) of the \HI\ compared to what
detected in large radio sources.

The \HI\ absorption in CSS/GPS shows often a relatively narrow component
(100-200 km/s FWHM), similar to what found against the nucleus of many
extended radio galaxies.  However, in addition to this, we have recently found
also cases of very broad \HI\ component.  Using the Westerbork Synthesis Radio
Telescope (WSRT) we have detected these very broad \HI\ absorption components
in a number of radio galaxies as reported in \citet{mo05a}.  An example of the
broad \HI\ absorption (for the radio galaxy 3C~305) is shown in Fig. 1 and
discussed below. The broad component is typically mostly blueshifted,
therefore indicating outflowing gas.  The objects showing this broad \HI\
component are characterised by the presence of a rich ISM surrounding the AGN,
e.g. strong CO or far-IR emission, and/or known to have undergone a major
episode of star formation in the recent past
\citep{ta05}.  Some of them have strong, steep-spectrum core
emission (on a scale $<$10 kpc, i.e. unresolved at the resolution of the WSRT
21-cm observations). These objects are considered to be young or recently
restarted radio sources (see e.g 3C~293 and 3C~236).

\begin{figure}
\centerline{\psfig{figure=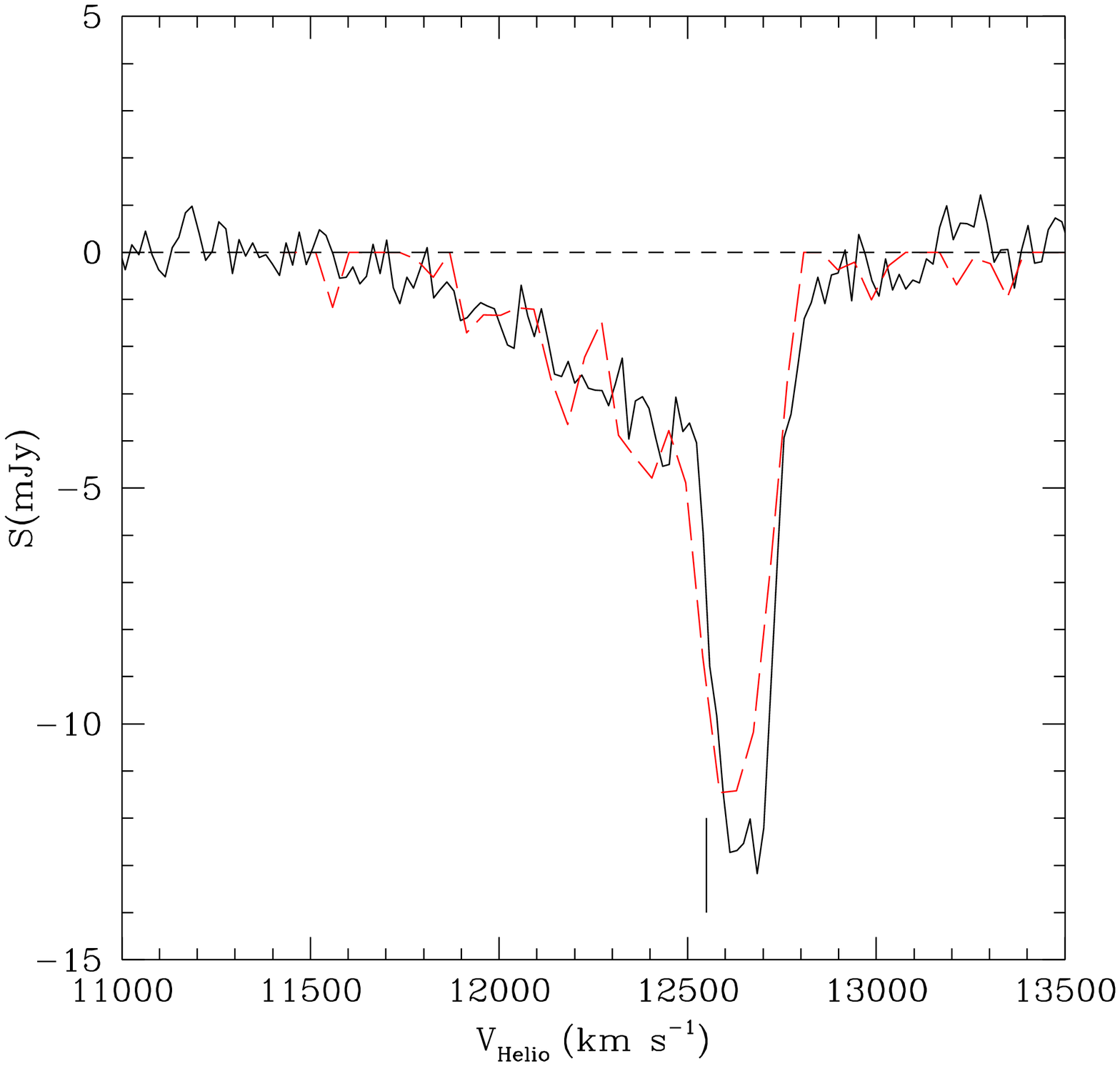,angle=0,width=5cm}
\psfig{figure=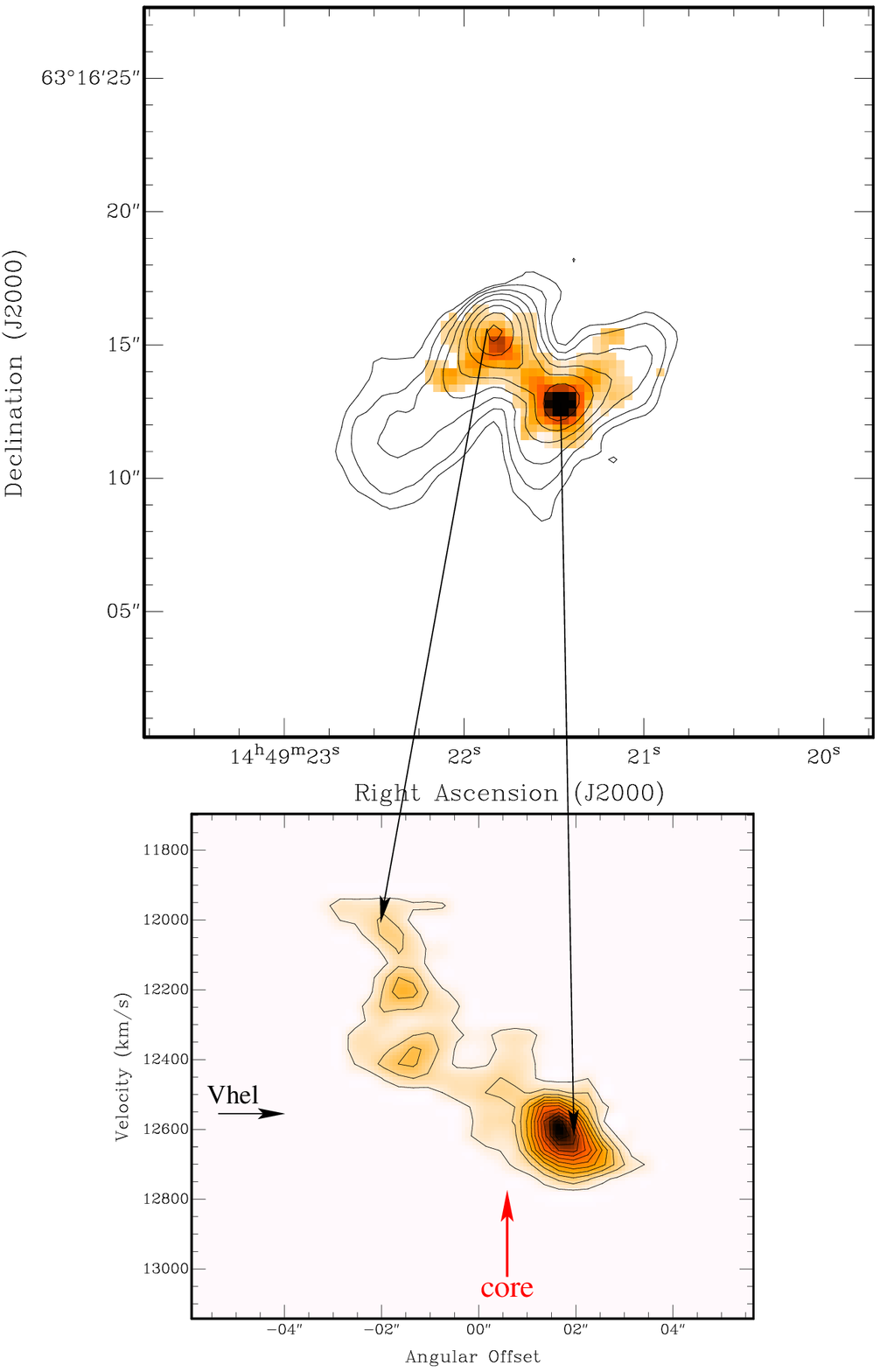,angle=0,width=5cm}
}
\caption{({\sl Left}) \HI\ absorption  profile obtained with the WSRT (solid
line) superimposed to the integrated spectrum from the - high spatial
resolution - VLA observations (long-dashed). The profile shows a deep,
relatively narrow, absorption and a broad component that covers more than 1000
\kms\ at zero intensity and mostly blueshifted compared to the systemic
velocity that is indicated.  ({\sl Right}) Panel showing {\sl (Top)} the radio
continuum image (contour) and the integrated \HI\ absorption (grey scale) from
the VLA data.  {\sl (Bottom)} The position-velocity plot from a slice passing
through the two lobes.  The grey scale image represents the total intensity of
the \HI\ absorption  \citep[see][for details]{mo05a}.}
\end{figure}

\subsection{The ionised gas}

Interestingly, in the same objects outflows with similar characteristics to the one in neutral
hydrogen have been detected  also in ionised gas.
\citet{ho05} and \citet{ho08} have studied the characteristics of the optical emission lines
in compact (GPS/CSS) and extended radio galaxies.  Of the 14 powerful CSS/GPS
studied (selected from the 3C/4C-2Jy samples), 11 show evidence for fast outflows
indicated by the presence of blueshifted components in the compact sources
more than in the extended.
The distributions of the velocities of the components are clearly different,
with the compact radio sources containing more extreme outflows than their
extended counterparts. Indeed, this trend is also evident within the sample of
compact radio sources, with the two highest outflow velocities observed in
some of the smallest (GPS) radio sources. The distributions were tested using
a Kolmogorov-Smirnoff test and were found to be different at the 99.9\%
confidence level. Hence, radio source size is clearly important in determining
the outflow velocity of the emission line gas \citep{ho07,ho08}.

\subsection{The location of the \HI\ outflow: the case of 3C~305 }

In order to understand whether the radio jet plays an important role in the
origin of the gas outflows described above, it is important to identify the location of the
outflows.  Because the radio sources we are interested in are small, this is
not an easy task.  We have now
this information for an handful of sources. In the case of the CSS 3C~305,
high-spatial resolution 21-cm \HI\ VLA observations have been obtained. These
high-resolution data show that the $\sim 1000$ \kms\ broad \HI\ absorption,
earlier detected in low-resolution WSRT observations, is occurring against the
bright, eastern radio lobe, about 1.6 kpc from the nucleus (see Fig. 1).
The broad \HI\ absorption has a  column density $2 \times
10^{21}$ cm$^{-2}$ (for T$_{spin}$=1000K). The mass outflowing gas is
estimated $\sim 10^6$ M$\sun$.

Earlier studies of the ionised gas had already found evidence for a strong
interaction between the radio jet and the interstellar medium at the location
of the eastern radio lobe \citep{jack03}. Our results show that the fast outflow produced by
this interaction also contains a component of neutral atomic hydrogen. The
most likely interpretation is that the radio jet ionises the ISM and
accelerates it to the high outflow velocities observed. Our observations
demonstrate that, following this strong jet-cloud interaction, not all gas
clouds are destroyed and that part of the gas can cool and become neutral
\citep[see also e.g.]{kr07}.

In addition to 3C~305 we have a few more objects where we know the gas outflow
(both neutral hydrogen and ionised gas) is happening off-nucleus. We have
interpreted all these outflows as due to {\sl interaction between that radio jet
and the ISM} \citep[see][for more details]{mo05b}.

\begin{figure}
\centerline{\psfig{figure=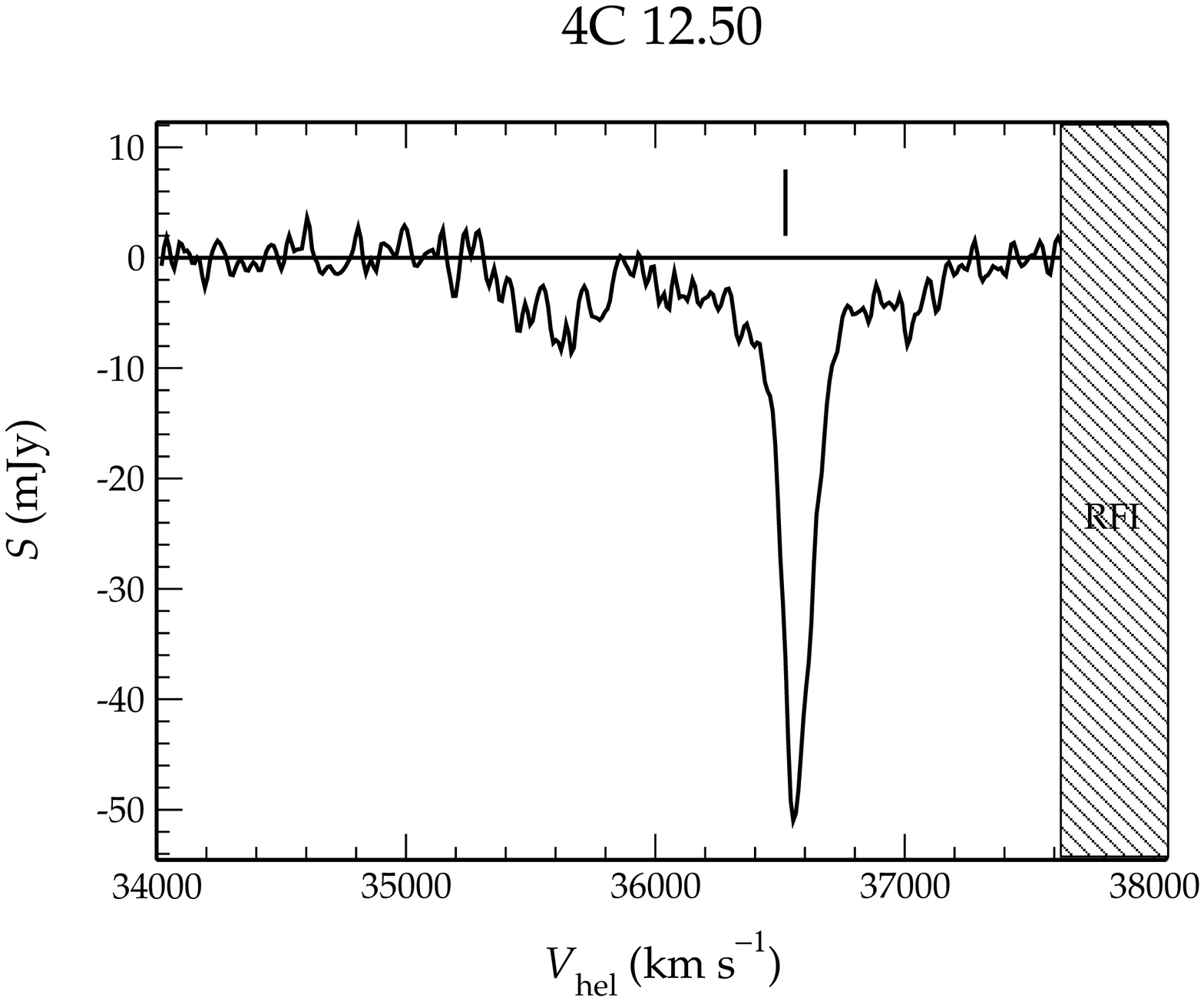,angle=0,width=6cm}
\psfig{figure=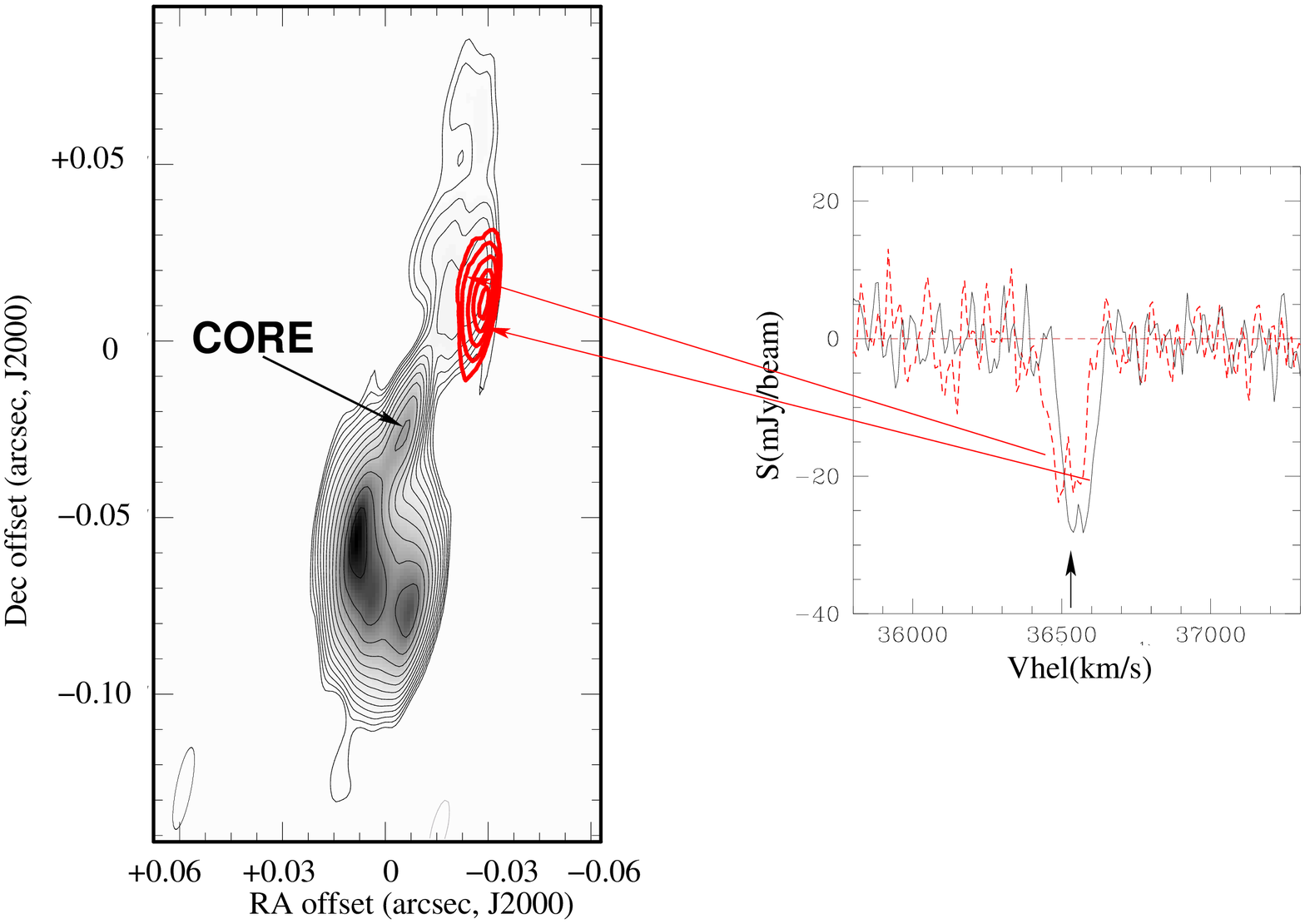,angle=0,width=8cm}
}
\caption{{\sl Left} \HI\ absorption profile of 4C12.50 as derived from WSRT
observations. The broad, blueshifted component is clearly seen. {\sl Right}
VLBI image of 4C12.50 and of the location of the narrower \HI\ absorption
component \citep{mo04}.  }
\end{figure}

\subsection{The mass outflow rate}

Mass outflow rates were calculated for all the galaxies where fast \HI\
outflows have been detected \citep[see][for details]{mo05a}. 

As mentioned above, the main result of this study is that the neutral outflows occur, in at least
some cases, at kpc distance from the nucleus, and they are most likely driven
by the interactions between the expanding radio jets and the gaseous medium
enshrouding the central regions.  We estimated that the associated mass outflow
rates are up to $\sim 50$ $M_\odot$ yr$^{-1}$, comparable (although at the
lower end of the distribution) to the outflow rates found for starburst-driven
superwinds in Ultra Luminous IR Galaxies (ULIRG), see \citet{ru02}.
This suggests that massive, jet-driven outflows of neutral gas in radio-loud
AGN can have as large an impact on the evolution of the host galaxies as the
outflows associated with starbursts. This is important as starburst-driven
winds are recognized to be responsible for inhibiting early star formation,
enriching the ICM with metals and heating the ISM/IGM medium.

\section{A detailed study of two objects}

\subsection{The extreme gas outflow in the GPS source 4C~12.50} 

As discussed in detail by \citet{ho03} and \citet{mo04}, PKS~1345+12 is
probably the best example of young radio galaxies showing extreme gas outflow
both in neutral hydrogen and ionised gas.  At the position of the nucleus we
observe complex emission line profiles and Gaussian fits to the [OIII]
emission lines require three components (narrow, intermediate and broad), the
broadest of which has width ~2000 \kms\ (FWHM) and is blueshifted by ~2000
\kms\ with respect to the halo of the galaxy and the narrow \HI\
absorption. We interpret this blueshifted component as material in outflow. We
also find evidence for large reddening [$0.92 <$E(B-V) $< 2.00$] and high
densities ($n_e> 4200$ cm$^{-3}$) for the most kinematically disturbed
component \citep[see][for details]{ho03}.

Interestingly a similarly
broad component is observed in \HI\ absorption (see Fig. 2).  The location of
the broad \HI\ outflow is not known yet. However, a VLBI study of the neutral
hydrogen in the nuclear regions of this object shows that the narrower \HI\
component (detected close to the systemic velocity) is associated with an
off-nuclear cloud ($\sim 50$ to 100 pc from the radio core, see Fig. 2) with a
column density of $\sim 10^{22}\ T_{\rm spin}/(100\ {\rm K}$) cm$^{-2}$ and an
\HI\ mass of a few times $10^5$ to $10^6$ M$_\odot$. A number of possibilities
have been considered to explain the results. In particular, this cloud appears
to indicate the presence of a rich and clumpy interstellar medium in the
centre, likely left over from the merger that triggered the activity and that
this medium influences the growth of the radio source.  The location of the
cloud -- at the edge of the northern radio jet/lobe -- suggests that the radio
jet might be interacting with this gas cloud.  This interaction could be
responsible for bending the young radio jet \citep[see][for details]{mo04}.
We argue that PKS~1345+12 is a young radio source with nuclear regions that
are enshrouded in a dense cocoon of gas and dust. The radio jets are expanding
through this cocoon, sweeping material out of the nuclear regions.

\begin{figure}
\centerline{\psfig{figure=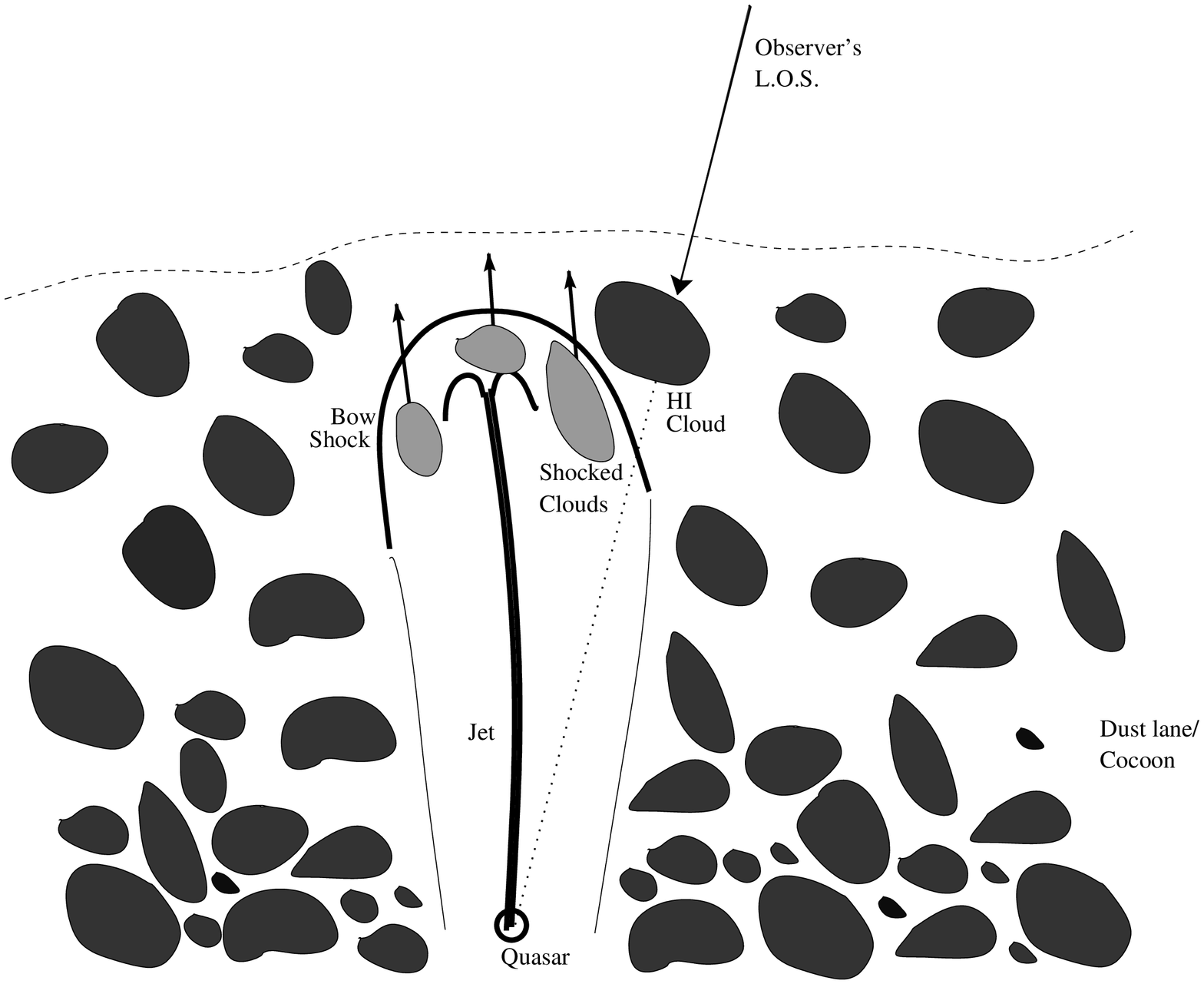,angle=-90,width=8cm}
\psfig{figure=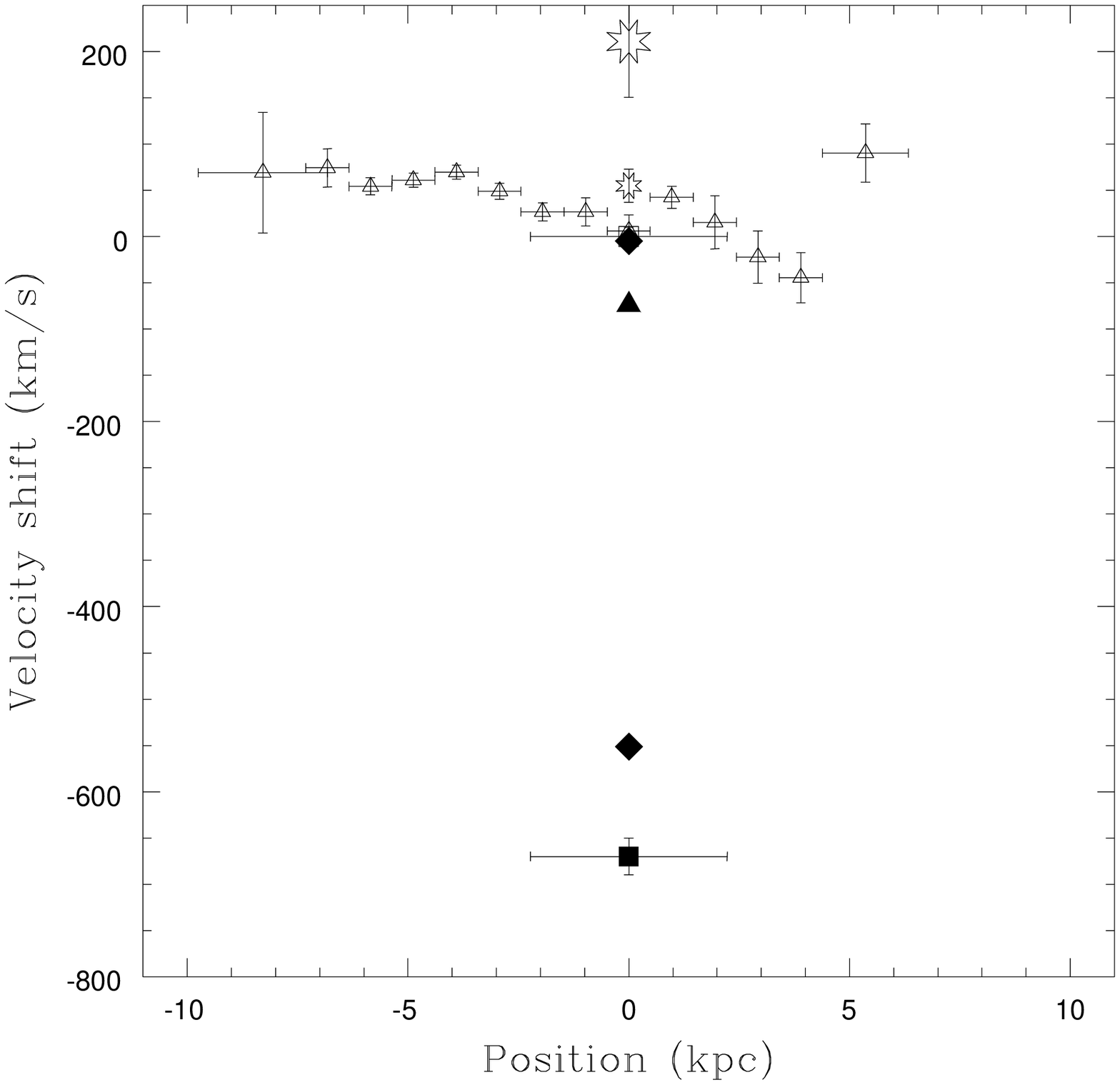,angle=0,width=6cm}}
\medskip
\centerline{\psfig{figure=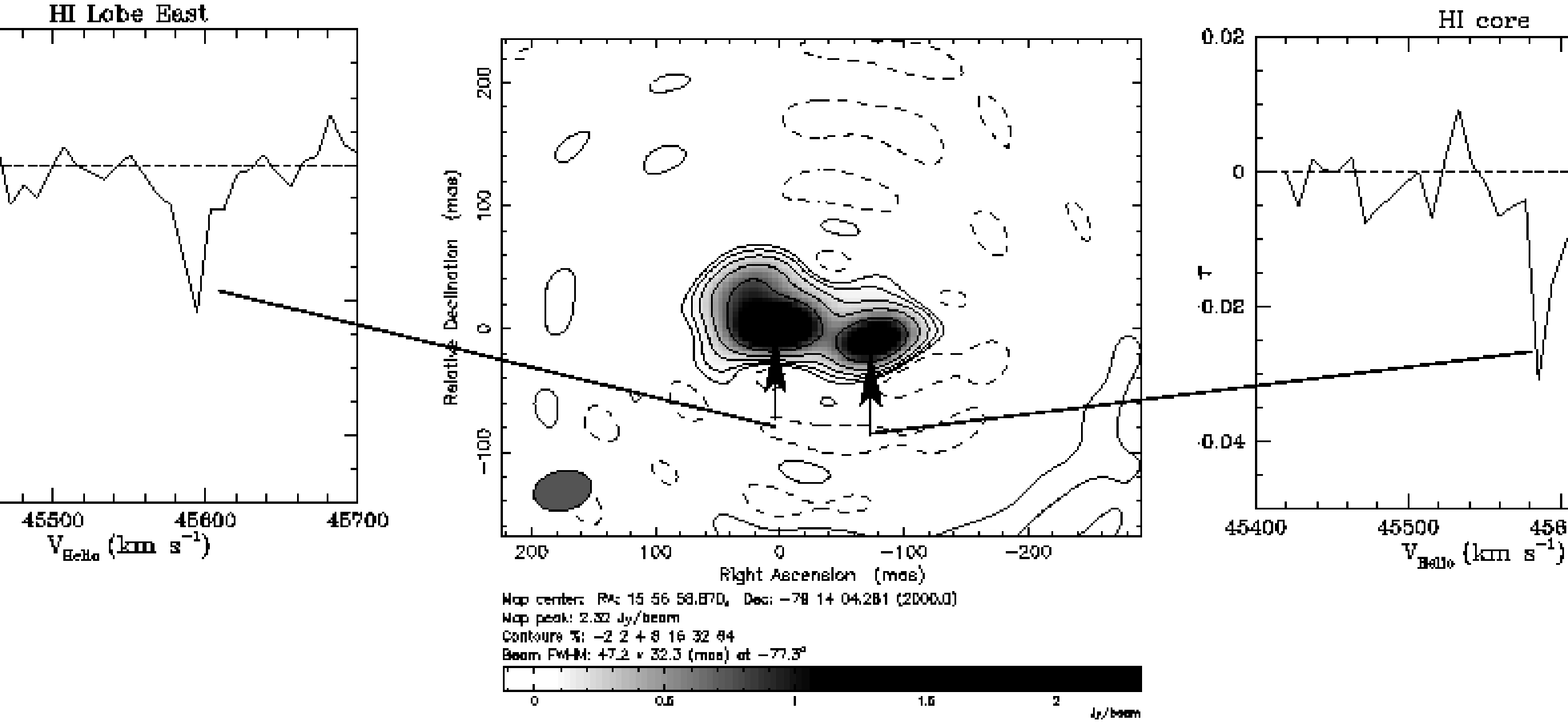,angle=-90,width=8cm}} 
\caption{
({\sl Top-left}) Cartoon representing the possible geometry arrangement for the
various emitting components in PKS~1549-79 \citep[see][for details]{ho06}. The [OIII]-emitting clouds are shaded light grey,
most of the [OII] is emitted by material in the extended disk/cocoon (shaded
light grey).   ({\sl Top-right}) Radial velocity profiles of the
optical emission lines of PKS 1549-79.  Small open and filled triangles
represent the narrow and intermediate components of H$\alpha$ respectively
\citep[symbols described in][]{ho06}.
Over-plotted is the radial velocity of the deep \HI\ 21cm absorption (large
filled triangle at -30~km s$^{-1}$ from \citet{mo01}).  ({\sl Bottom})
\HI\ absorption from the VLBI data of PKS~1549-79. This shows as the narrow
\HI\ component is undisturbed gas that surrounds the entire radio source \citep[see][for details]{ho06}.}
\end{figure}

\subsection{PKS~1549-79: an example of radio source in the early-stage of its
evolution}

The compact radio source PKS~1549-79 provides the best example that in the
initial growth phase, the black hole is substantially obscured at
optical wavelengths in this object, consistent with recent galaxy evolution
models \citep{ho05}.  We use deep optical, infrared and radio observations to
explore the symbiosis between nuclear activity and galaxy evolution in this
compact radio source \citep[see][and ref. therein for more details]{ho06}.
The optical images reveal the presence of tidal tail features
which provide strong evidence that the host galaxy has undergone a major
merger in the recent past. The merger hypothesis is further supported by the
detection of a young stellar population (YSP), which, on the basis of spectral
synthesis modeling , was formed 50-250 Myr ago and makes up a significant
fraction of the total stellar mass (1-30 per cent).

Despite the core-jet structure of the radio source, which is consistent with
the idea that the jet is pointing close to our line of sight, our \HI\ 21-cm
observations reveal significant \HI\ absorption associated with both the core
and the jet, see Fig.3 and \citet{ho06}.
No broad permitted (optical) lines are detected  but broad Pa$\alpha$ in NIR
is observed. The optical lines (e.g. [O III]\AA 5007,4959 lines) are  highly blueshifted 
($\Delta V \sim 680$ \kms) indicating the presence of a fast outflow.

We conclude that PKS 1549-79 is a radio source in a stage where the nucleus is
still hidden (in the optical) by the gas/dust coming from the merger that
triggered the radio source.  Young, small scale radio jets are expanding
through dense cocoon sweep aside gas and dust.  The AGN driven outflow will
eventually remove this gas.  However, we find that the {\sl warm-gas outflow}
is small (mass outflow rates $0.12 < \dot{M} < 12$~M$_{\odot}$ yr$^{-1}$) and not as large as expected in the quasars feedback model
\citep[see][for details]{ho07}. We are now investigating whether outflows of hot or
cold gas may have a more significant impact.

\section{Nearby compact radio galaxies}

The objects described above are all powerful compact radio galaxies.  For
nearby ($z < 0.04$) lower luminosity, compact sources the situation could be
different.  It is interesting to look at the characteristics of very nearby
compact steep spectrum as the one discussed by Giroletti these Proceedings.

Some of these objects are so close that \HI\ in emission can be detected and
studies.  This was done as part of the observations of a large sample of
nearby radio galaxies \citep{em06}.
The remarkable trend found is that the radio galaxies with large amounts
(M$_{\rm \HI} >10^9 \ M_\odot$) of
extended (many tens of kpc up to 200 kpc!) \HI\ disks all have a compact radio
source while such disks have not be found in extended radio sources (in
particular in FR type-I radio galaxies). Fig. 4 shows an example of such huge
\HI\ structures and the plot illustrating how the very \HI-rich structures are
associated with sub-kpc radio galaxies.

This trend can be explained if the \HI-rich compact radio sources do not grow
into extended sources, either because frustrated by the ISM in the central
region of the galaxy or because the fuel stops before the source expands.
This would therefore support the idea (see also Giroletti these proceedings)
that not all compact sources are young counterparts of the extended radio
galaxies.

\begin{figure}
\centerline{\psfig{figure=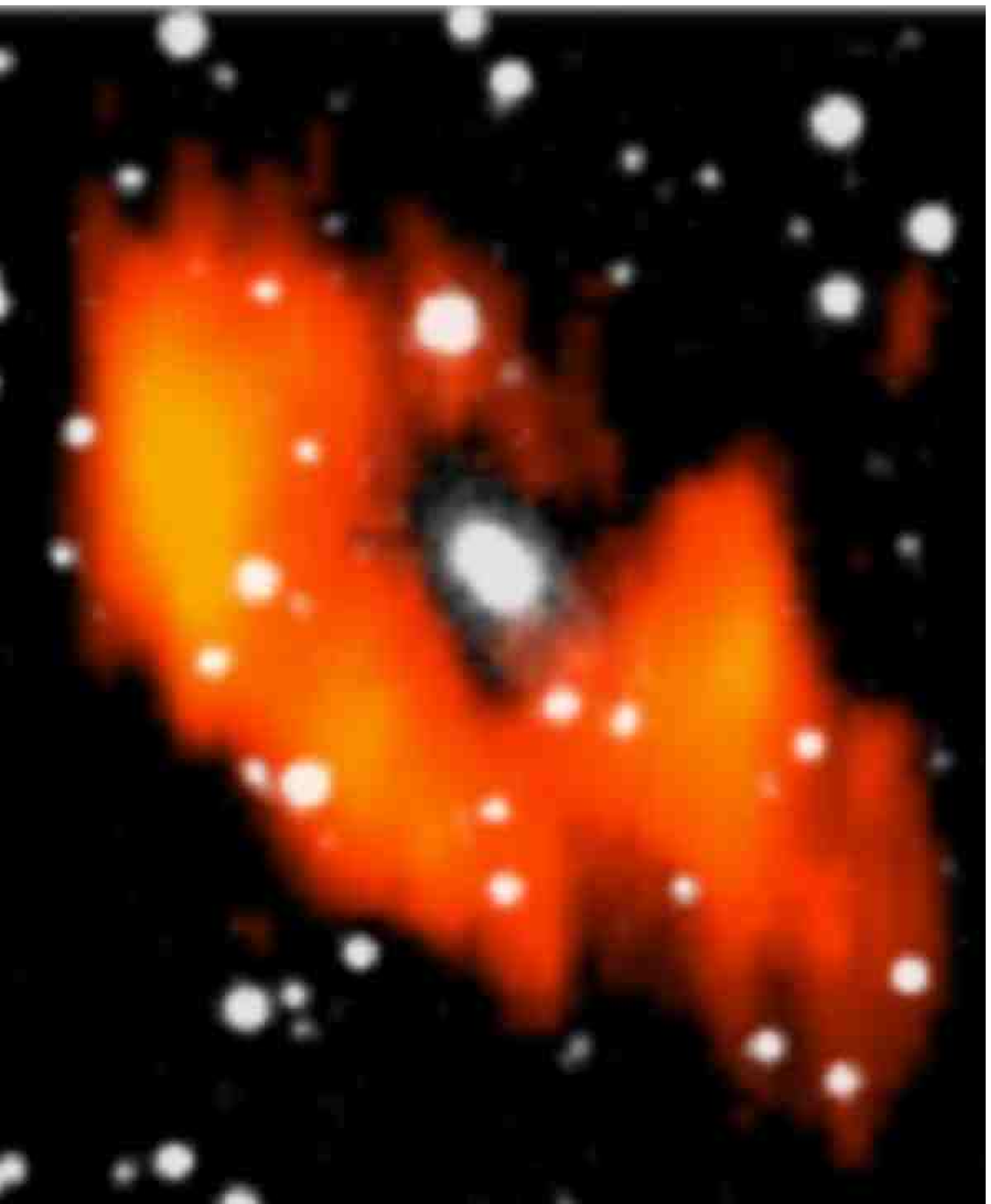,angle=0,width=6cm}
\psfig{figure=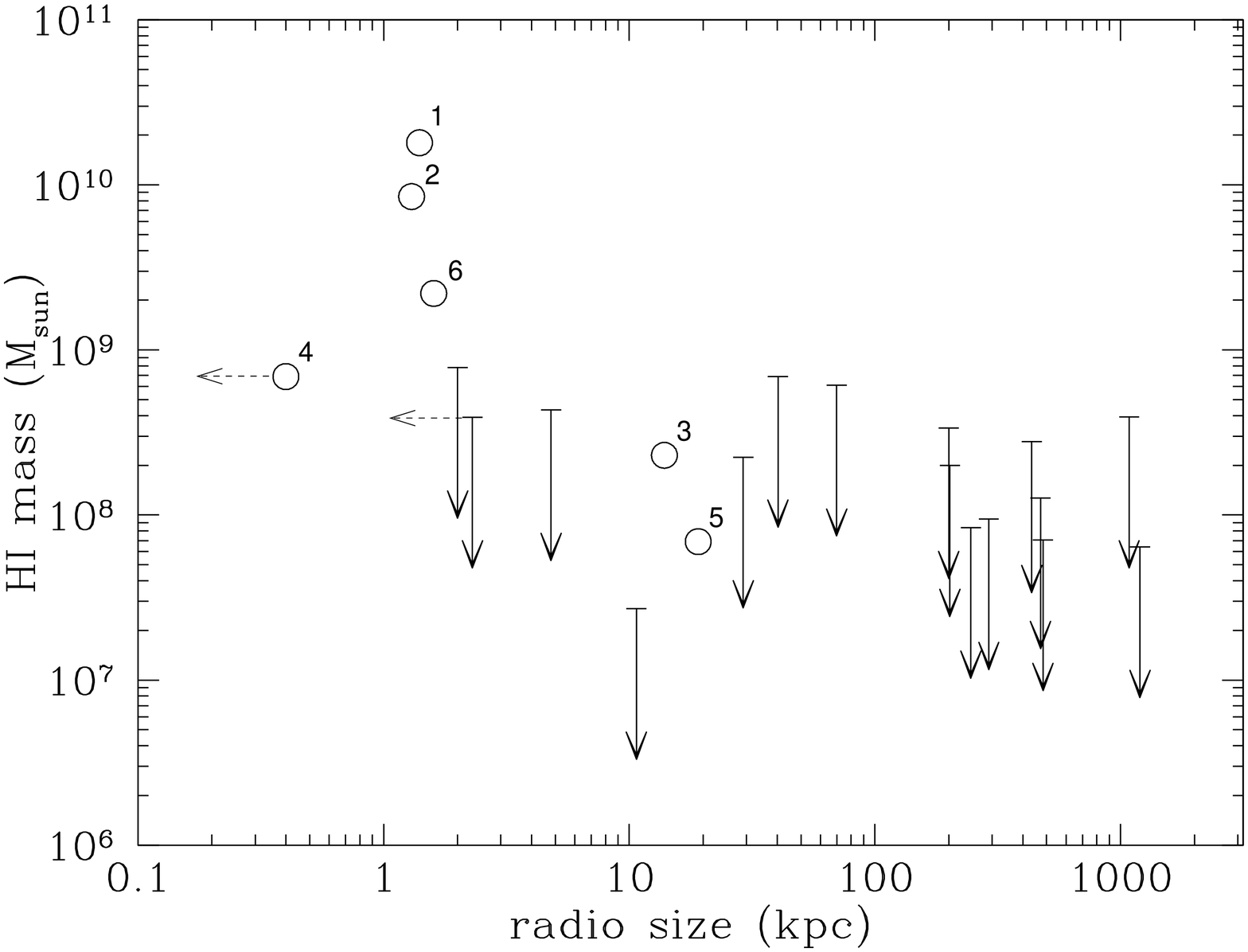,angle=0,width=8cm}
}
\caption{{\sl Left} Total \HI\ intensity (orange in the electronic version of
the paper) superimposed to the optical image of B2 0648+25 . {\sl Right} Plot of
the HI mass versus the size of the radio source: it shows as large amount of
\HI\ (mainly distributed in large disk structure) were detected only in
compact kpc-scale sources. Figures taken from \citep{em06}}
\end{figure}

\section {Summary of the results}

The results presented here support the idea that a rich medium is often found
around young radio galaxies.  Broader and blueshifted optical emission lines
are associated with such sources indicating that fast outflow of ionised gas
are very common.  High \HI\ column density and fast \HI\ outflows are detected
and interpreted as produced by the interaction between the radio jet and the
surrounding dense medium.  Thus, in powerful compact sources the jet/ISM
interaction is important and many effects are seen, including complex, stratified
structure of the ionised gas outflow.  The mass of the gas involved in the
outflows does not seem to be large enough to frustrate the source but likely
able of 
slowing down the evolution of the jets.
Nearby low-luminosity compact sources may have a different evolution. Some of
them may just fade away and do not grow to large size.


\acknowledgements 

I would like to thank the organizing committee and in particular Travis Rector
and Dave de Young  for organizing and inviting me at this very pleasant and
interesting workshop.  The results presented in this review would not have
been obtained without the help of my collaborators. In particular I would like
to thank Tom Oosterloo, Clive Tadhunter, Joanna Holt and  Bjorn Emonts.


\end{document}